\documentclass{article}
\usepackage{spconf,amsmath,graphicx}

\usepackage{amsmath}
\usepackage{amssymb}
\usepackage{booktabs}
\usepackage{array}
\usepackage{multirow}
\usepackage{tabularx} 
\usepackage{verbatim}

\usepackage[breaklinks,colorlinks]{hyperref}

\usepackage[capitalize]{cleveref}
\crefname{section}{Sec.}{Secs.}
\Crefname{section}{Section}{Sections}
\Crefname{table}{Table}{Tables}
\crefname{table}{Tab.}{Tabs.}
\usepackage{gensymb}

\usepackage{subcaption}

\usepackage{fancyhdr} 

\fancypagestyle{firstpage}{
  \fancyhf{} 
  \fancyhead[L]{Submitted to IEEE ICIP 2024} 
  \fancyfoot{} 
}

\newcommand{\etal}{\textit{et al.} }

\title{SALIENCY-AWARE END-TO-END LEARNED VARIABLE-BITRATE \\ 360-DEGREE IMAGE COMPRESSION}
\name{Oğuzhan Güngördü, A. Murat Tekalp}
\address{Department of Electrical and Electronics Engineering, Koc University, 34450 Istanbul, Turkey}
\begin{document}
\ninept
\maketitle
\thispagestyle{firstpage} 
\begin{abstract}
Effective compression of 360\textdegree{} images, also referred to as omnidirectional images (ODIs), is of high interest for various virtual reality (VR) and related applications. 2D image compression methods ignore the equator-biased nature of ODIs and fail to address oversampling near the poles, leading to inefficient compression when applied to ODI. We present a new learned saliency-aware 360\textdegree{} image compression architecture that prioritizes bit allocation to more significant regions, considering the unique properties of ODIs. By assigning fewer bits to less important regions, significant data size reduction can be achieved while maintaining high visual quality in the significant regions. To the best of our knowledge, this is the first study that proposes an end-to-end variable-rate model to compress 360\textdegree{} images leveraging saliency information. The results show significant bit-rate savings over the state-of-the-art learned and traditional ODI compression methods at similar perceptual visual quality. Supplementary materials are available at \href{https://sigport.org/documents/saliency-aware-end-end-learned-variable-bitrate-360-degree-image-compression-supplementary}{\textit{[Supplementary URL]}}.
\end{abstract}
\section{Introduction}
\label{sec:intro}
Unlike conventional 2D images, omnidirectional images (ODIs) capture a scene with a comprehensive 360\textdegree{} $\times$ 180\textdegree{} field of view (FoV), encompassing every angle and direction. This panoramic view provides a more immersive experience, allowing viewers to explore the scene in any direction.
However, this capability comes with some challenges for storage, streaming, and rendering of ODI. A primary concern is the inherently higher resolution of ODIs compared to standard 2D images. Efficient compression of 360\textdegree{} images is of paramount importance, as it leads to substantial bandwidth savings, enhanced energy efficiency, and reduced delays in many applications. ODIs also have oversampling issues near the poles, causing non-uniform resolution, making objects near the poles appear larger and those near the equator smaller. Addressing these challenges is essential for efficient ODI compression, viewport prediction, and image quality assessment \cite{litreview360}. 

The visual information captured by ODIs is vast; however, not all regions of ODIs garner equal attention. Certain areas near the latitude or those containing salient objects, tend to attract more fixations, while others might remain largely overlooked \cite{Sitzmann_TVCG_VR-saliency}. This uneven distribution of viewer attention provides distinct opportunities to address efficient 360\textdegree{} image compression by combining the tasks of saliency detection and 2D image compression. 

In conventional 2D images, viewers often focus at the center, anticipating it to contain salient information \cite{centerbias}. In contrast, in the spherical domain of ODIs, attention predominantly lies around the equator. Studies reveal that viewers don't fully explore ODIs; even doubling the viewing time from 10s to 20s doesn't always yield new fixations \cite{lookaroundyou}. Thus, certain areas of ODI are frequently ignored. Based on these observations, this paper shows saliency prediction systems addressing variations in visual attention to highlight areas that are likely to attract human focus can be used to improve the overall compression efficiency of ODI. 

To this effect, we first introduce an end-to-end learned model for saliency prediction specifically tailored to compression of ODIs. Our saliency model is designed to capture both global and local visual attention information, effectively addressing the unique characteristics of 360\textdegree{} image content.
We then combine this saliency information with the state-of-the-art (SOTA) 2D learned image compression to guide the bit allocation by latent space masking, ensuring that salient regions are preserved with higher fidelity. Our key contributions are summarized as follows: 
\begin{itemize}
    \item We propose a new end-to-end learned dual-stream saliency model specifically tailored to the compression of ODIs by integrating four sub-networks: one sub-network generates the Attention stream capturing global saliency, the second generates the Expert stream focusing on local saliency, the third is the Coordinate Refinement network that refines both streams for equator bias, and the Fusion network merges dual refined streams. Our model achieves SOTA performance in terms of correlation coefficient (CC) and is competitive in other metrics without any imbalance on the Salient360! \cite{salient360_2017} and Saliency in VR \cite{Sitzmann_TVCG_VR-saliency} datasets. \vspace{-2.5pt} 
    \item We propose extending learned 2D image compression (LIC) architectures to saliency-aware ODI compression by incorporating ODI saliency maps into 2D LIC architectures via latent masking, adeptly prioritizing both informative regions and non-oversampled areas, especially latitudes, due to the equator bias in 360\textdegree{} images projected by equirectangular projection (ERP). The proposed approach guides bit allocation in a learnable manner, using the Sal-MSE distortion loss emphasizing salient regions during training to mitigate the oversampling challenge inherent in 360\textdegree{} images.  \vspace{-2.5pt} 
    \item The proposed scheme is a variable-rate ODI compression model, eliminating the need for separate training at different bit rates. To the best of our knowledge, this is the first end-to-end variable-rate 360\textdegree{} image compression model leveraging saliency information. \vspace{-2.5pt} 
    \item Our approach outperforms SOTA learned and traditional conventional 2D and 360\textdegree{} image codecs in terms of WS-PSNR, SAL-PSNR, and overall visual quality. 
\end{itemize}

\section{Related Work}
\label{sec:format}

\subsection{360\textdegree{} Image Saliency Detection}

Rising interest in 360\textdegree{} image content has necessitated the development of saliency prediction techniques specialized for this medium. Among hand-crafted methods, Startsev \etal\cite{STARTSEV201843} addressed projection distortions and equator bias in 360\textdegree{} images by averaging three 2D saliency predictors. Battisti \etal\cite{battisti} developed a model using both low and high-level image features, adjusting the averaged saliency map with an equator-focused weighting map. Lebreton \etal\cite{LEBRETON201869} adapted traditional 2D saliency models for the omnidirectional realm, introducing variants such as GBVS360 and BMS360.

More closely related to our work are deep learning based saliency prediction models specifically designed for 360\textdegree{} content. Monroy \etal\cite{monroy2018salnet360} introduced SalNet360, which integrates spherical coordinates into a CNN for equator-bias learning. 
MV-SalGAN360 \cite{9122430} extended SalGAN360 \cite{8551543} by projecting 360\textdegree{} images into 2D views 
, then fusing them using a weighted sum based on each view's FoV. MRGAN360 \cite{gao2023mrgan360} used recurrent networks for inter-stage correlations, and ATSal \cite{dahou2020atsal} implemented a dual-stream architecture, merging outputs through pixel-wise multiplication.

Different from the literature reviewed above, we propose a dual stream model to capture global and local saliency maps separately, compensate them for equator bias, and then fuse these two saliency maps with a lightweight CNN. Our architecture offers flexibility as each component can be individually adjusted. Crucially, we bypass the typical equator-prior weighting and linear averaging, enhancing robustness and eliminating extra pre and post-processing steps that require dataset-specific adjustments.

\subsection{360\textdegree{} Image Compression}
360\degree{} image compression is essential for the growing VR applications due to the high resolution of omnidirectional content, challenging storage, and transmission. The ERP's latitude-dependent sampling density makes conventional 2D image codecs suboptimal for 360\textdegree{} content in terms of both complexity and performance. Hadizadeh and Bajić \etal\cite{saliency_aware_video} incorporated visual attention into their HEVC-based saliency-aware video codec by adjusting QP according to saliency to address the complexity challenge. Jiang \etal\cite{hevc_perceptual_2d_compression} similarly included a spatiotemporal saliency model into H.265/HEVC-based codecs. For 360\textdegree{} content, Chiang \etal\cite{saliency_driven_rdo_360} and Luz \etal\cite{8122228} both introduced HEVC-based solutions using machine learning saliency detection and adaptive QP strategies.

Most closely related to our work are the learned compression architectures specific to 360\textdegree{} images. LIC360 \cite{lic360} pioneered an end-to-end model for 360\textdegree{} images in ERP format. However, it was primarily tested on low-resolution images with extended viewport trajectories and CMPs, missing out on ERP metrics such as WS-PSNR and SAL-PSNR. Its latitude adaptive scale network solely relies on height, and its code estimation network often allocates more bits to generic regions.  Not considering the code mask in bit-per-pixel calculations can affect performance when using their method for the saliency-aware architectures. OSLO \cite{mahmoudian2022oslo} modifies learned 2D architectures for the spherical domain. While efficient for the spherical domain, its output isn't directly viewable in the ERP domain, posing practical limitations. The lack of a comprehensive 360\textdegree{} dataset limits the training of such architectures, especially when a saliency map and a 360\textdegree{} image are both essential. OSLO's emphasis on sphere learning also misses the benefits of fine-tuning with large 2D image datasets.  

Attention mechanisms in 2D image compression have been underscored in \cite{liang2023sigvic, roi_comp}, introducing variable rates and attention in a learnable way, a feature not present in 360\textdegree{} architectures. \cite{saliency_perceptual_comp} integrated masking into learned 2D codecs, but their binary saliency map assumption, manual weighted loss, and lack of a variable rate mechanism limit their approach. In contrast, our method embeds a saliency mechanism as a latent mask in the compression architecture, leveraging the equator bias of 360\textdegree{} saliency maps to prioritize informative areas. Contrary to the dual systems in \cite{lic360}, we assign fewer bits to over-sampled regions by learnable masking with the saliency output.

\section{Saliency-Aware ODI Compression}
\label{sec:majhead_arch}
\subsection{Saliency Detection Architecture}
\label{ssec:sal_arch_sec}

\begin{figure*}
  \centering
  \includegraphics[width=\linewidth]{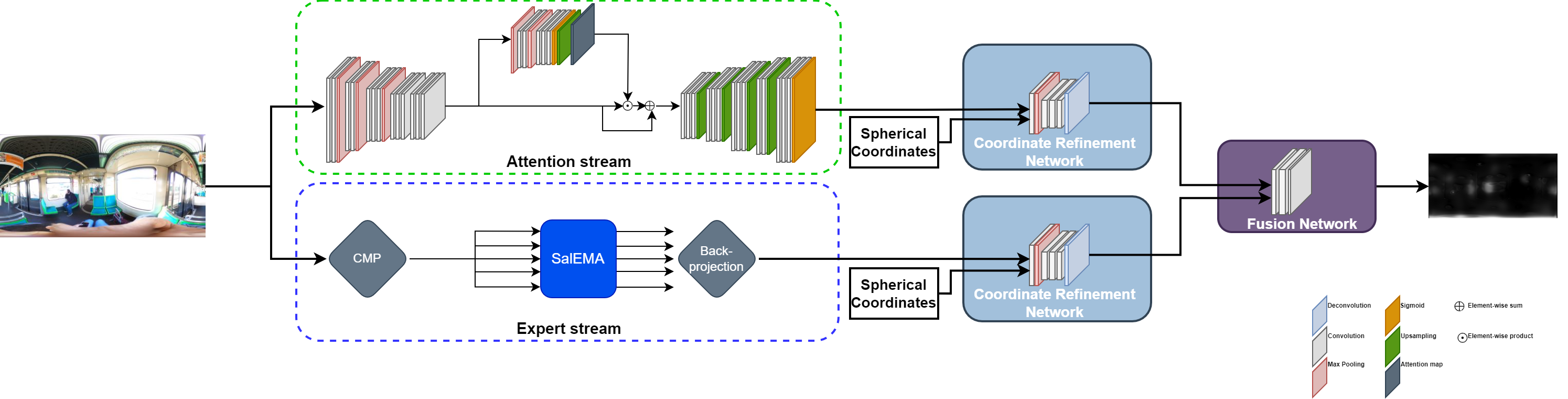}
  \caption{The dual-stream architecture of the proposed end-to-end ODI saliency learning model. The Attention Stream learns global saliency, while the Expert Stream learns local saliency using cube map patches. The Coordinate Refinement Network with Spherical Coordinates learns to compensate for the equator bias, and the Fusion Network combines two streams to generate the final saliency map.}
  \label{fig:saliency_architecture}
\end{figure*} 

The proposed saliency detection framework that is tailored to ODI compression addresses four main challenges: detection of local and global saliency masks, compensating both masks for equator bias, learning to fuse global and local saliency masks, and ensuring stable performance across different datasets without fine-tuning hyperparameters. The proposed saliency detection architecture, depicted in \cref{fig:saliency_architecture}, consists of two streams: the upper (Attention) stream captures global saliency, while the lower (Expert) stream focuses on local details using cube map patches processed by SalEMA \cite{SalEMA}. We then concatenate spherical coordinates to the saliency map, feeding it into the Coordinate Refinement network that learns equator bias adjustment. Finally, the Fusion network seamlessly combines global and local saliency information.

Unlike ATSal \cite{dahou2020atsal}, which uses different networks for poles and equator cube map patches, our uniform application of SalEMA \cite{SalEMA} across all patches cuts computational costs and simplifies training. Additionally, we apply coordinate refinement directly to the predicted ERP saliency maps after the dual stream, in contrast to SalNet360 \cite{monroy2018salnet360}, which does so for six cube map patches. This reduces the computational load from six operations to two, efficiently compensating for the equator bias. The outputs of Attention and Expert streams are concatenated with 2-channel per pixel spherical coordinates that denote latitude and longitude information. This concatenated map is then fed into the Coordinate Refinement network to learn the equator bias. This network consists of a sequence of convolutional layers with the number of filters and their sizes as follows: ({32}×\(5 \times 5\)), max-pooling of \(2 \times 2\), ({64}×\(3 \times 3\)), ({64}×\(5 \times 5\)), ({32}×\(5 \times 5\)), ({1}×\(7 \times 7\)), and a transposed convolutional layer ({1}×\(4 \times 4\)). ReLU activation function is applied after each layer.

Utilizing a Fusion network, our model generates the final saliency map by merging saliency information from two distinct streams, diverging from \cite{zhang2019video}'s application in spatiotemporal contexts. Our approach ensures a precise and unified saliency representation facilitated by a unique integration method and a distinct loss function, as detailed in \cref{sec:loss_functions}. The fusion process is carried out through a two-layer convolutional network ({64, 128}×\(3 \times 3\)) to merge the two saliency maps coming from parallel streams, followed by a (\(1 \times 3 \times 3\)) convolutional network to yield a single map. This design transforms our saliency architecture into a lightweight, end-to-end deep learning model, autonomously learning and adapting without the need for manually imposed local or global saliency information or equator bias through averaging. 

\subsection{Saliency-aware Compression Architecture}
\label{ssec:comp_arch_sec}

\begin{figure}
  \centering
  \includegraphics[width=\linewidth]{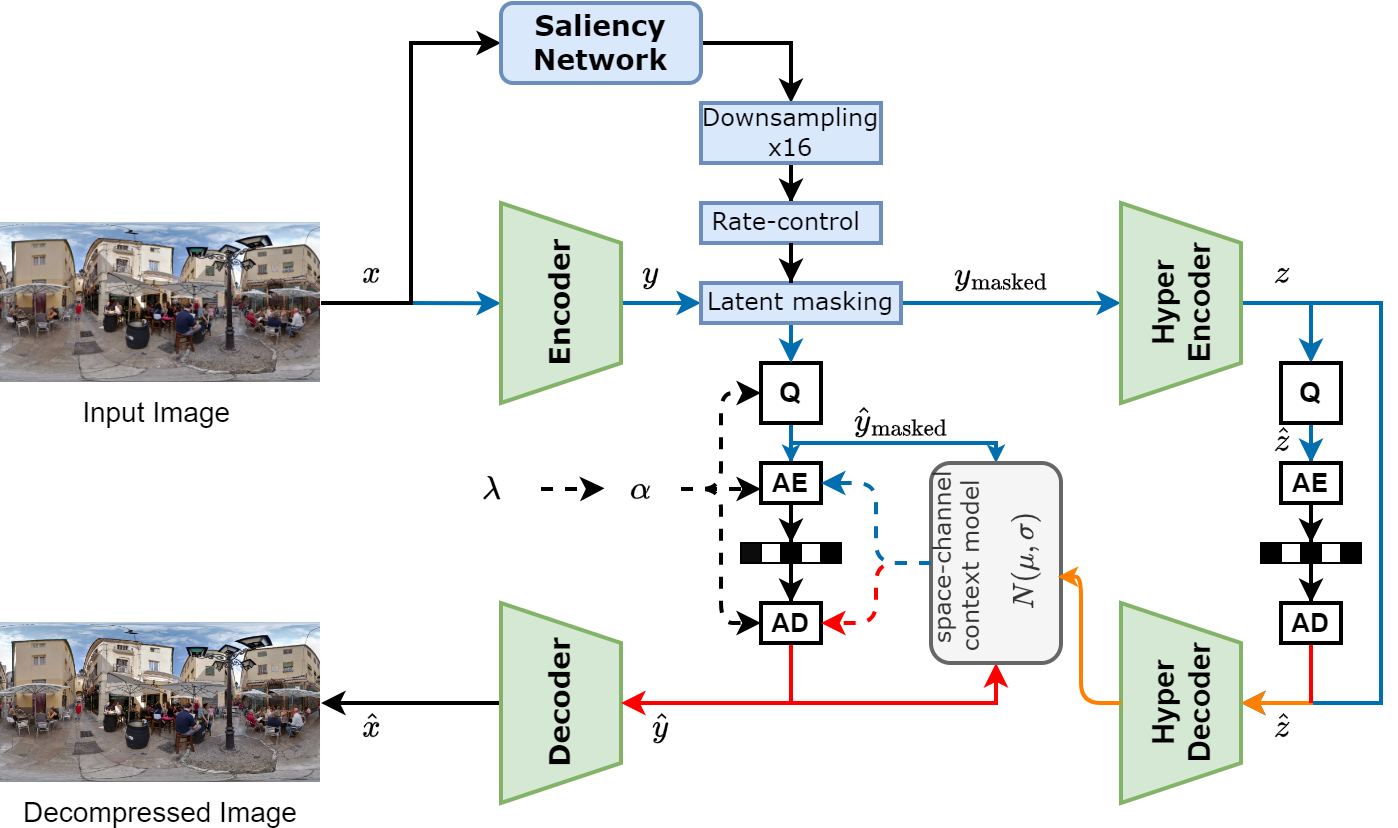}
  \caption{Block diagram of the saliency-aware ODI compression architecture, where blue boxes indicate proposed modifications to a given LIC model shown by green boxes. Blue/red arrows denote data flow for encoding/decoding, and orange is shared by both.}
  \label{fig:compression_architecture}
\end{figure}

We describe an ODI compression model that leverages the saliency information inherent in 360\degree{} images to enhance the perceptual compression performance. 
The core concept entails weighting the latent representation by a saliency mask. This strategy aims to retain more information in the salient regions while allowing for more aggressive compression in the non-salient regions. The image compression model described here is inspired by the Efficient Learned Image Compression (ELIC) architecture \cite{elic_comp} although the proposed latent space masking concept can be applied to other learned compression architectures as well. We note that a different form of latent masking has been previously applied to the region of interest coding in \cite{roi_comp}.

The process starts by predicting the saliency map of the input image \( x \) using our saliency network shown in \cref{fig:saliency_architecture}. Unlike \cite{roi_comp}, we rescale the predicted saliency map between 0 and 255 before applying a sigmoid activation. This rescaling is crucial as the proportion of salient regions to non-salient regions is relatively low compared to 2D images. We aim to preserve the value differences between salient and non-salient regions, which would be diminished by smoothing operations or by applying a sigmoid function to a saliency map in the range of 0 to 1. The downsampling block, as depicted in \cref{fig:compression_architecture}, is employed to perform a 16x downscale operation on the saliency map to match the dimensions of the latent space, as expressed in the following equation:
\begin{equation}
\scalebox{1.1}{$\text{Saliency Mask} = \text{AvgPool}(\sigma(\text{Saliency Network}(x)))$}
\end{equation}
Next, to control rate allocation, we first obtain the Saliency Mask Residual using the following equation, setting \(\alpha = 1\) in the rate control block:
\begin{equation}
\scalebox{1.1}{$\text{Saliency Mask Residual} = \left( \text{Saliency Mask} + \alpha \right) / \alpha$}
\label{eq:saliency_mask_residual}
\end{equation}
Subsequently, we multiply this Saliency Mask Residual with the latent space to enforce the saliency map in the latent masking operation:
\begin{equation}
\scalebox{1.1}{$y_{\text{masked}} = y_{\text{ch}0-\text{ch}47} \, || \, (y_{\text{ch}48-\text{ch}191} \odot \text{\small Saliency Mask Residual})$}
\label{eq:modified_latent_space}
\end{equation}
In the latent representation \( y \) from the compression network, which has 192 channels, the first 48 feature maps are preserved. The subsequent channels are weighted using \cref{eq:saliency_mask_residual} to incorporate the saliency map into the latent space, as shown in \cref{eq:modified_latent_space}. By preserving the initial 48 channels, we ensure sufficient information for the background is retained. This not only prevents the fading of its reconstructed texture but also ensures competitive performance in weighted-to-spherically-uniform-PSNR (WS-PSNR) rather than focusing solely on Saliency-PSNR (SAL-PSNR).

Latent masking ensures higher fidelity preservation of the salient regions during the compression and decompression processes. The architecture is trained end-to-end to minimize a loss function that balances the reconstruction error in the salient regions and the bits required to represent the compressed image. Through this saliency-aware approach, our architecture aims to achieve a more perceptually pleasing compression, particularly focusing on retaining the quality of salient regions in the images. Unlike the approach in \cite{lic360}, our compression architecture only utilizes the saliency map at the encoder side, eliminating the need to send the saliency map to the decoder side, which would increase the bits per pixel (bpp) significantly.

\subsubsection{Variable Rate Compression Model} 
\label{sssec:var_rate_sec}
In our quest for a variable rate ODI compression model, we're inspired by the Quantization-error-aware Variable Rate Framework (QVRF) introduced by Tong \etal\cite{tong2023qvrf}. Three-stage training ensures optimal performance across diverse bit rates without individual training for each. Initially, the network is optimized at a fixed Lagrange multiplier \( \lambda = 0.18 \), setting a foundation for minimized distortion. The second stage jointly refines the quantization regulator vector \( \mathbf{A} \), controlling the overall quantization error of the latent representation, and noise approximation, using a predefined set of Lagrange multipliers \( \mathbf{\lambda} = (0.0018, 0.0035, 0.0067, 0.013, 0.025, 0.0483, 0.0932, 0.18) \). The final stage employs straight-through estimation to fine-tune both the network and \( \mathbf{A} \), ensuring adaptability to varied bit rates while maintaining quality.

\subsection{Training Loss Functions}
\label{sec:loss_functions}
In this section, we elucidate the loss functions utilized for training both the saliency and compression components of our architecture. The Attention and Expert streams are trained using the loss function as defined in the ATSal \cite{dahou2020atsal} and SalEMA \cite{SalEMA}, respectively. The Coordinate Refinement network employs an L2 loss, while the Fusion network's loss incorporates both Kullback-Leibler Divergence (KLD) and Correlation Coefficient (CC) losses. The Fusion network loss, \( \mathcal{L}_{\text{FusionNet}} \), is formulated as:
\begin{equation}
\scalebox{1.2}{$\mathcal{L}_{\text{FusionNet}} = \mathcal{L}_{\text{KLD}} - \mathcal{L}_{\text{CC}}$}
\end{equation}

On the compression side, our architecture employs a modified Mean Squared Error (MSE) distortion loss, which we refer to as Sal-MSE loss, to train the compression model. Given a raw image \( \mathbf{x} \) and reconstructed image \( \hat{\mathbf{x}} \) obtained through our compression architecture, the compression loss is formulated as follows:
\begin{equation}
\scalebox{1.2}{$\mathcal{L} = \lambda \cdot \mathcal{L}_{\text{Sal-MSE}} + \mathcal{L}_{\text{bpp}}$}
\end{equation}
where \( \lambda \) is a trade-off between rate and distortion, \( \mathcal{L}_{\text{Sal-MSE}} \) is the masked Mean Squared Error loss, and \( \mathcal{L}_{\text{bpp}} \) is the bits-per-pixel loss. This loss function aims to minimize the reconstruction error in the salient regions of the image while also minimizing the bits required to represent the compressed image. The Sal-MSE loss is computed as:

\begin{equation}
\scalebox{1.2}{$\mathcal{L}_{\text{Sal-MSE}} = \frac{1}{\sum S} \sum_{i=1}^{N} S_i \cdot \left( \mathbf{x}_i - \hat{\mathbf{x}}_i \right)^2$}
\end{equation}
where \( N \) is the total number of pixels in the image, \( S \) is the sigmoid of the ground truth saliency map, and the sum in the denominator is taken over all values in the \( S \). 

\section{Experiments}
\subsection{Experimental Setup}
\label{sec:experimental}

\subsubsection{Training Details} Our saliency and compression architectures are trained separately, allowing future integration of advanced saliency models without affecting compression. Both are first fine-tuned with SALICON 2D images, then with Salient360! 360\degree\ images to account for the unique resolution distribution and object size perception in omnidirectional images. To overcome the scarcity of extensive 360\degree\ datasets, we use mirroring and horizontal flipping on \cite{salient360_2017} and \cite{Sitzmann_TVCG_VR-saliency}.

Our saliency architecture undergoes two-stage training, with all images and maps scaled to a resolution of $640\times320$. The Attention and Expert streams, inspired by ATSal \cite{dahou2020atsal} and SalEMA \cite{SalEMA}, start with pre-trained weights. Initially, only the Attention, Expert, and Fusion networks are trained using SALICON images, excluding the Coordinate Refinement network due to the absence of equator bias in 2D images. In the next stage, all sub-networks are trained concurrently on 360\degree\ images, end-to-end, based on their specific loss functions. The network is trained over 100 epochs, using a batch size of 4, a learning rate of $1.3e-06$, and the Adam optimizer with a ReduceLROnPlateau strategy, having a factor of 0.5 and a patience of 10 epochs.

The compression architecture undergoes training in three stages tailored to the variable rate model. Each stage begins with training on 2D images, followed by fine-tuning using augmented 360\degree\ images. Images and saliency maps are scaled to resolutions of $512\times512$ and $2048\times1024$, respectively, ensuring the retention of oversampling artifacts. By employing random crops of $1024\times512$ and $512\times1024$ from 360\degree\ images, we capture differences in equator and pole sampling. This strategy ensures each batch contains a mix of both regions, mitigating the risk of overfitting specifically to one region. Each stage starts with a learning rate of $1e-04$, a batch size of 16, and 200,000 iterations on 2D images. For the 360\degree\ image fine-tuning, the learning rate is reduced to $1e-05$, the batch size to 8, with an added 10,000 iterations.

\subsubsection{Evaluation} Our saliency and compression architectures were assessed using Salient360! test images at their original resolution, aligning with the models submitted to the Salient360! Grand Challenge ICME2017 benchmark \cite{salient360_2017}. Additionally, the saliency model was tested on the Saliency in VR dataset \cite{Sitzmann_TVCG_VR-saliency} to observe differences in detecting visually salient areas. The best result of each experiment is shown in bold in the following subsections. For saliency detection, four metrics from the Salient360! dataset \cite{salient360_2017} were employed: Normalized Scanpath Saliency (NSS), Kullback-Leibler Divergence (KLD), Linear Correlation Coefficient (CC), and AUC-Judd (AUC-J). Compression architectures were evaluated using three 360\degree\ ERP-specific metrics: WS-PSNR \cite{Sun2017WeightedtoSphericallyUniformQE}, SAL-PSNR \cite{8122228}, and WS-SSIM \cite{8652269}. WS-PSNR weighs pixel inaccuracies based on latitude; SAL-PSNR emphasizes errors in salient regions and near equator latitudes; and WS-SSIM considers structural differences in 360\degree\ images, accounting for the ERP.


\begin{figure}
    \centering
    \small 
    \begin{tabular}{@{}c ccc}
        \raisebox{1.5\height}{\textbf{SalNet360}} & \includegraphics[width=0.22\linewidth]{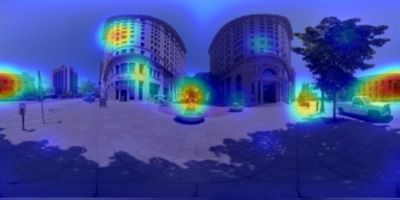} & \hspace{-4mm} \includegraphics[width=0.22\linewidth]{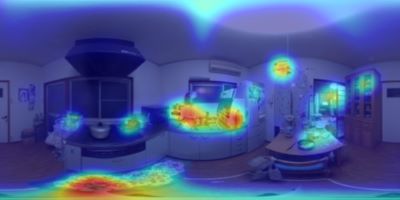} & \hspace{-4mm} \includegraphics[width=0.22\linewidth]{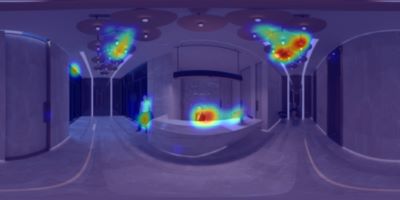}  \\
        \raisebox{1.5\height}{\textbf{MV-SalGAN360}} & \includegraphics[width=0.22\linewidth]{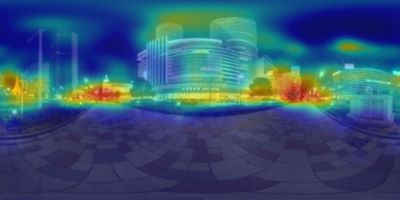} & \hspace{-4mm} \includegraphics[width=0.22\linewidth]{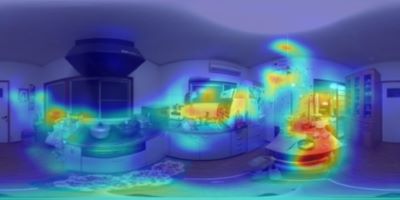} & \hspace{-4mm} \includegraphics[width=0.22\linewidth]{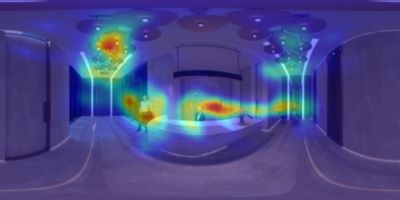} \\
        \raisebox{1.5\height}{\textbf{MRGAN360}} & \includegraphics[width=0.22\linewidth]{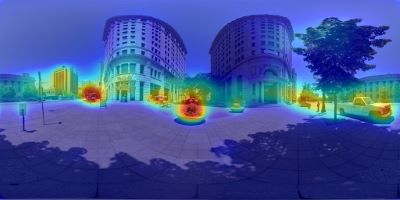} & \hspace{-4mm} \includegraphics[width=0.22\linewidth]{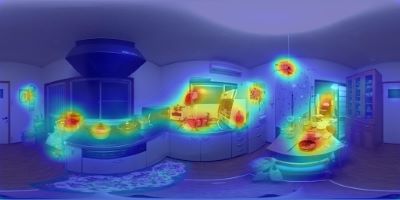} & \hspace{-4mm} \includegraphics[width=0.22\linewidth]{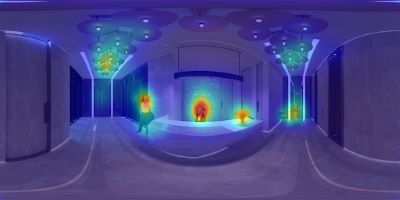} \\
        \raisebox{1.5\height}{\textbf{ATSal}} & \includegraphics[width=0.22\linewidth]{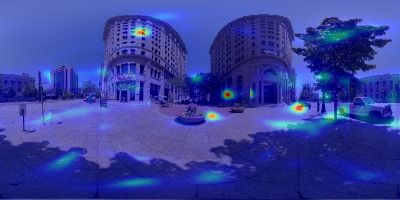} & \hspace{-4mm} \includegraphics[width=0.22\linewidth]{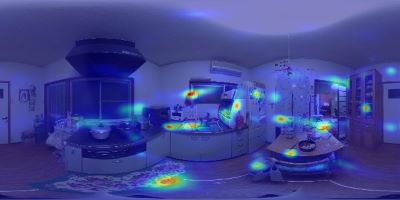} & \hspace{-4mm} \includegraphics[width=0.22\linewidth]{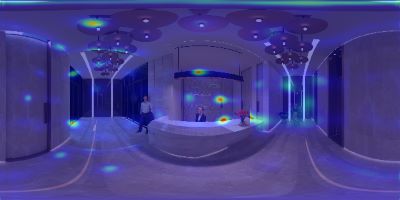} \\
        \raisebox{1.5\height}{\textbf{Ours}} & \includegraphics[width=0.22\linewidth]{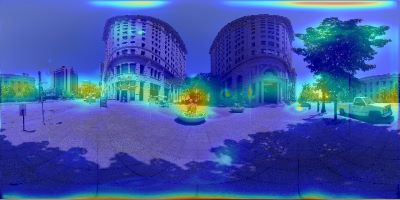} & \hspace{-4mm} \includegraphics[width=0.22\linewidth]{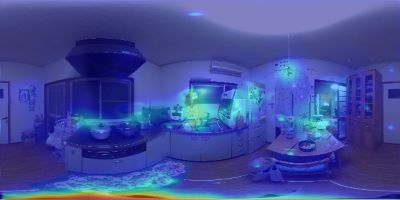} & \hspace{-4mm} \includegraphics[width=0.22\linewidth]{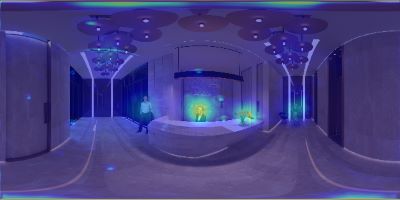} \\
        \raisebox{1.5\height}{\textbf{Ground Truth}} & \includegraphics[width=0.22\linewidth]{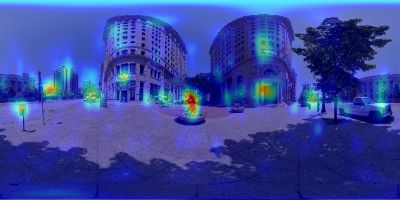} & \hspace{-4mm} \includegraphics[width=0.22\linewidth]{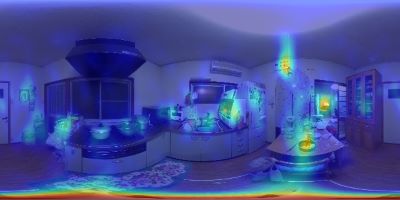} & \hspace{-4mm} \includegraphics[width=0.22\linewidth]{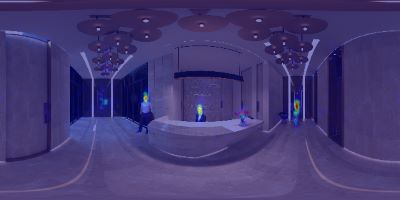} \\
    \end{tabular}
    \caption{Visual comparison with other 2D and 360\degree{} saliency models: the first two columns represent the Salient360! 2017 test set \cite{salient360_2017}, and the last column the Saliency in VR test set \cite{Sitzmann_TVCG_VR-saliency}.}
    \label{fig:visual_comparison_sal}
\end{figure}

\subsection{Comparison with the SOTA: Saliency Detection}

Our proposed saliency architecture's performance is visually benchmarked against four models in \cref{fig:visual_comparison_sal}. This set includes four 360\degree\ models: SalNet360 \cite{monroy2018salnet360}, MV-SalGAN360 \cite{9122430}, ATSal \cite{dahou2020atsal}, and MRGAN360 \cite{gao2023mrgan360}, chosen based on available source code. Benchmark results from the Salient 360! Grand Challenge ICME2017 \cite{salient360_2017} are detailed in \cref{tab:saliency_metric_comparison}. Our model excels in the CC metric, while MRGAN360 stands out in NSS and AUC-J, and MV-SalGAN360 in KLD. However, our model remains competitive across all metrics.

The integration of Fusion and Coordinate Refinement networks in our architecture enhances saliency detection at poles, surpassing the models we were inspired by. Our end-to-end approach, without manual saliency map combination, ensures consistent results across various datasets. While MV-SalGAN360 \cite{9122430} is recognized as one of the best saliency models, it often misses finer details our model captures, especially in the Saliency in VR \cite{Sitzmann_TVCG_VR-saliency}. Conversely, 2D models like BMS, when applied to 360\degree\ images, tend to over-label regions as salient.

\begin{table}
  \centering
  \small
  \begin{tabular}{c c c c c}
    \toprule
    Method & KLD $\downarrow$ & CC $\uparrow$ & NSS $\uparrow$ & AUC-J $\uparrow$\\
    \midrule
    SalNet360 \cite{monroy2018salnet360} & 0.458 & 0.548 & 0.755 & 0.701\\
    SalGAN \cite{pan2018salgan} & 1.236 & 0.452 & 0.810 & 0.708\\
    GBVS360 \cite{LEBRETON201869} & 0.698 & 0.527 & 0.851 & 0.714\\
    BMS360 \cite{LEBRETON201869} & 0.599 & 0.554 & 0.936 & 0.736\\
    SalGAN\&FSM \cite{7965634} & 0.896 & 0.512 & 0.910 & 0.723\\
    SalGAN360 \cite{8551543} & 0.431 & 0.659 & 0.971 & 0.746\\
    MV-SalGAN360 \cite{9122430} & \textbf{0.363} & 0.662 & 0.978 & 0.747\\
    MRGAN360 \cite{gao2023mrgan360} & 0.401 & 0.658 & \textbf{1.09} & \textbf{0.784}\\
    SalBiNet360 \cite{salbinet360} & 0.402 & 0.661 & 0.975 & 0.746\\
    ATSal \cite{dahou2020atsal} & 0.449 & 0.630 & 0.865 & 0.693\\
    Ours & 0.406 & \textbf{0.669} & 0.981 & 0.737\\
    \bottomrule
  \end{tabular}
  \caption{Comparison of saliency detection methods on the test set of Salient360! 2017 Benchmark. \cite{salient360_2017}.}
  \label{tab:saliency_metric_comparison}
\end{table}

\subsection{Comparison with the SOTA: ODI Compression}
\subsubsection{RD Performance} 
\begin{figure}
  \vspace{-10pt}
  \centering
  \begin{subfigure}{0.23\textwidth}
    \includegraphics[width=\linewidth]{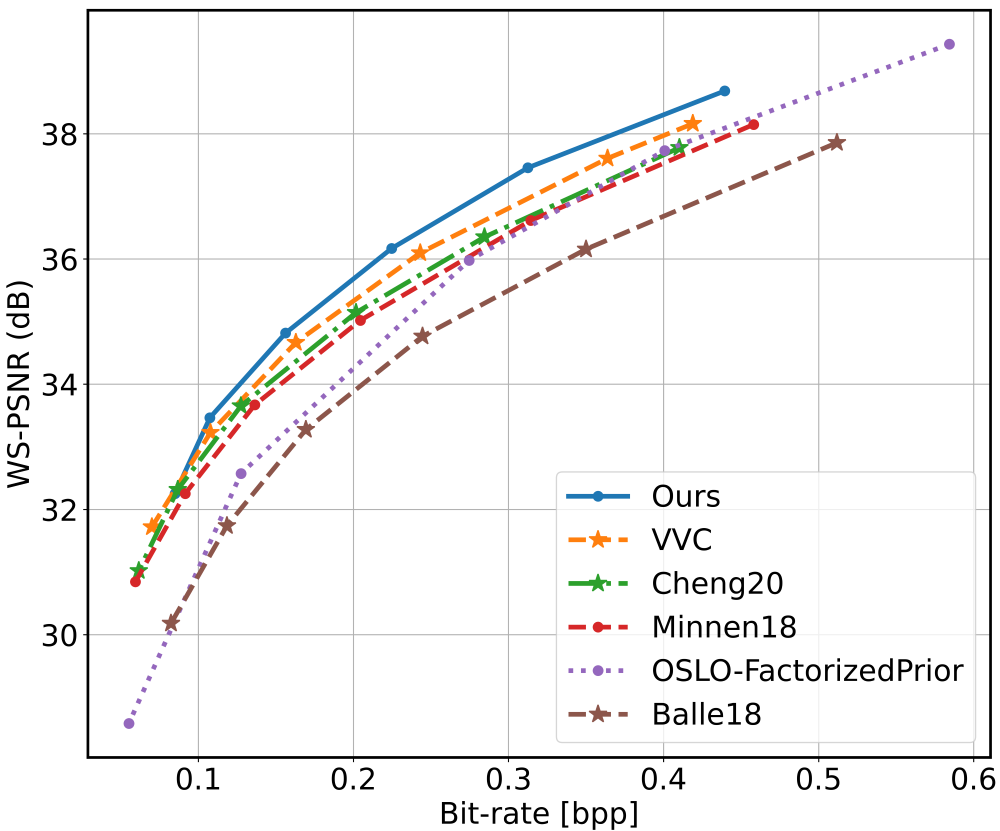}
  \end{subfigure}
  \begin{subfigure}{0.23\textwidth}
    \includegraphics[width=\linewidth]{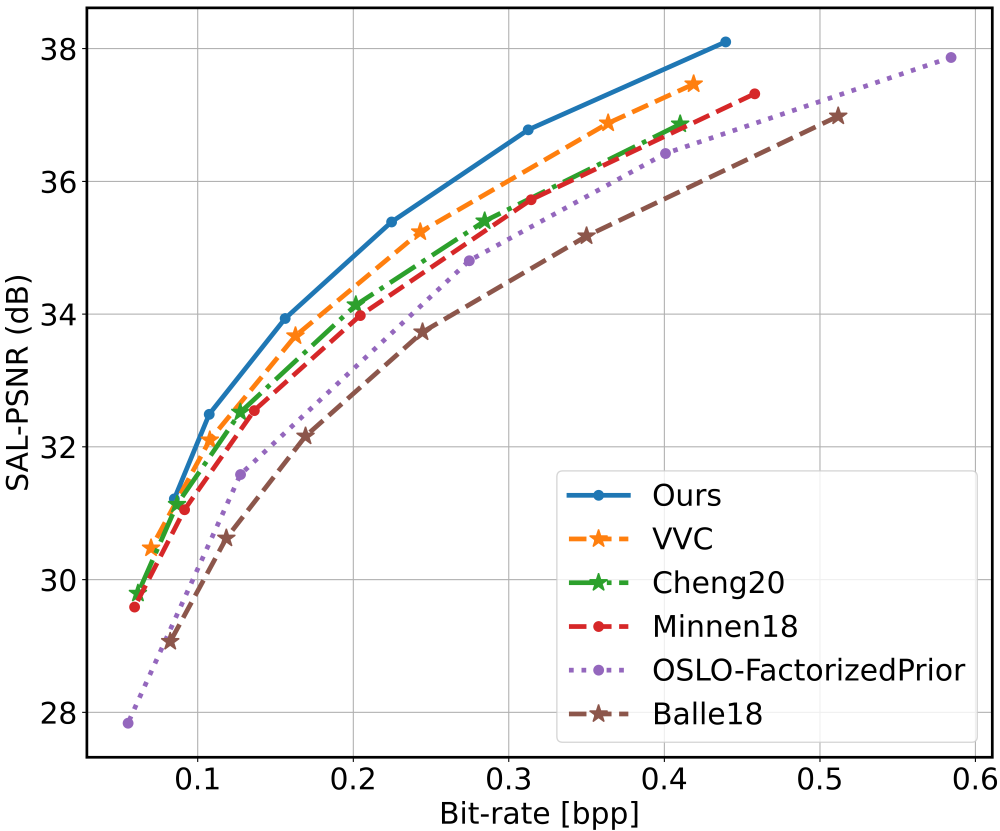}
  \end{subfigure}
  \begin{subfigure}{0.23\textwidth}
    \includegraphics[width=\linewidth]{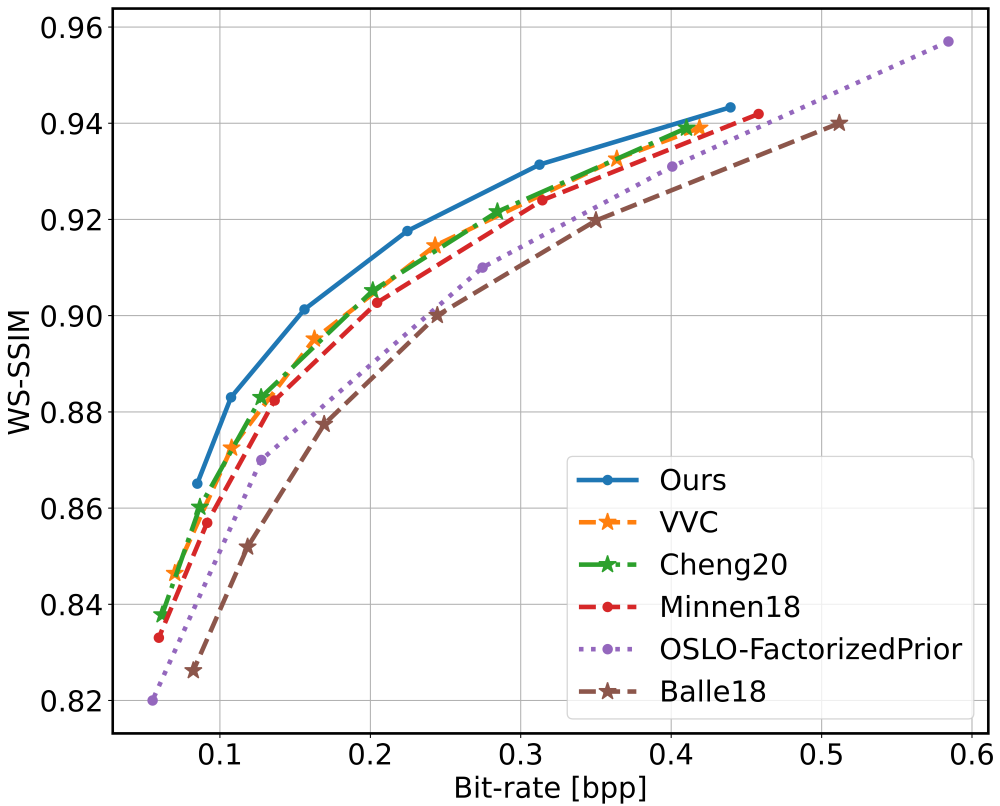}
  \end{subfigure}
  \caption{WS-PSNR, SAL-PSNR, and WS-SSIM RD curves for different compression methods on Salient 360! 2017 test set\cite{salient360_2017}.}
  \label{fig:rd_plots}
  \vspace{-10pt}
\end{figure}

We compare the RD performances of our compression architecture with other learned compression methods \cite{ballé2018variational,minnen2018joint,cheng2020learned,lic360,mahmoudian2022oslo} optimized by MSE and the standard VVC VTM\cite{vvc2021} in \cref{fig:rd_plots}. Our architecture outperforms other methods in terms of SAL-PSNR and WS-SSIM on Salient 360! test images. Regarding WS-PSNR, our model is slightly better than the VVC VTM \cite{vvc2021} and achieves significantly better coding performance than other learned codecs by preserving the first 48 channels in the latent space as recommended in \cite{roi_comp}. We also present Bjontegaard delta (BD) \cite{bjontegaard2001calculation} analysis computed from WS-PSNR and SAL-PSNR curves, with the reference RD performance anchored to Ball\'{e}2018 \cite{ballé2018variational}. As shown in \cref{tab:bd_results}, in the case of SAL-PSNR, our method achieves BD-PSNR gain of 1.93 dB and BD-rate savings of 36.60\%, which outperform other methods, even with the incorporation of a variable-rate mechanism.

\begin{table}
    \centering
    \small
    \setlength{\tabcolsep}{2.5pt} 
    \begin{tabular}{c c c|c c}
        \hline
        \multirow{2}{*}{Method} & \multicolumn{2}{c|}{WS-PSNR} & \multicolumn{2}{c}{SAL-PSNR} \\
        & BD-psnr & BD-rate & BD-psnr & BD-rate \\
        \hline
        Ball\'{e}18 \cite{ballé2018variational} & 0 dB & 0\% & 0 dB & 0\% \\
        Minnen18-mean \cite{minnen2018joint} & 0.59 dB & -13.78\% & 0.58 dB & -12.97\% \\
        Minnen18 \cite{minnen2018joint} & 1.13 dB & -25.60\% & 1.16 dB & -25.24\% \\
        Cheng20 \cite{cheng2020learned} & 1.32 dB & -28.93\% & 1.35 dB & -28.51\% \\
        VVC-VTM (yuv444) \cite{vvc2021} & 1.56 dB & -32.27\% & 1.68 dB & -33.17\%  \\ \hline
        OSLO-FactorPrior \cite{mahmoudian2022oslo} & 0.67 dB & -13.91\% & 0.61 dB & -13.45\% \\
        LIC360 \cite{lic360} & 0.74 dB & -14.52\% & 0.63 dB & -14.01\% \\
        Ours & \textbf{1.69 dB} & \textbf{-34.01\%} & \textbf{1.93 dB} & \textbf{-36.60\%} \\
        \hline
    \end{tabular}
    \caption{Comparison of BD-PSNR and BD-Rate vs. Ball\'{e}2018 \cite{ballé2018variational} on Salient 360! 2017 test set \cite{salient360_2017}. Methods above mid line are 2D image codecs. Methods below mid line are ODI codecs.}
    \label{tab:bd_results}
\end{table}

\subsubsection{Visual Quality} 
\begin{figure*}
    \centering
    \begin{subfigure}{0.226\textwidth}
        \includegraphics[width=1.8cm]{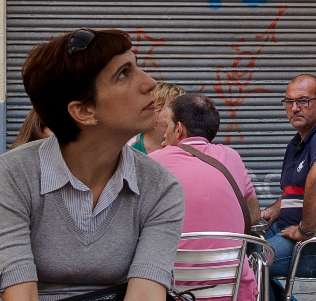}
        \includegraphics[width=1.8cm]{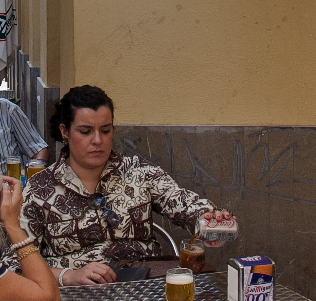}
        \caption*{Original crops}
    \end{subfigure}
    \hspace{0.005cm} 
    \begin{subfigure}{0.226\textwidth}
        \includegraphics[width=1.8cm]{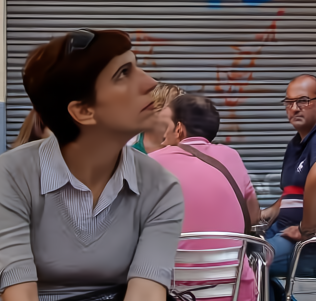}
        \includegraphics[width=1.8cm]{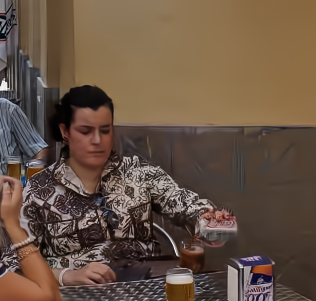}
        \caption*{Ours (0.145/30.75/0.843)}
    \end{subfigure}
    \hspace{0.005cm} 
    \begin{subfigure}{0.226\textwidth}
        \includegraphics[width=1.8cm]{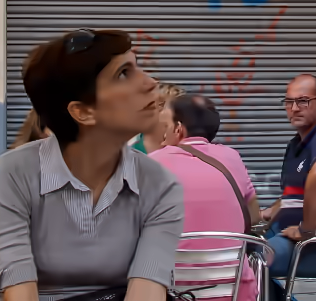}
        \includegraphics[width=1.8cm]{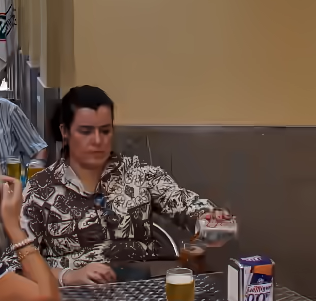}
        \caption*{VVC (0.132/30.27/0.825)}
    \end{subfigure}
    \hspace{0.005cm} 
    \begin{subfigure}{0.226\textwidth}
        \includegraphics[width=1.8cm]{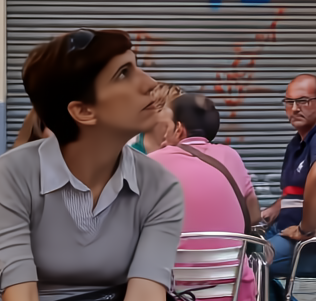}
        \includegraphics[width=1.8cm]{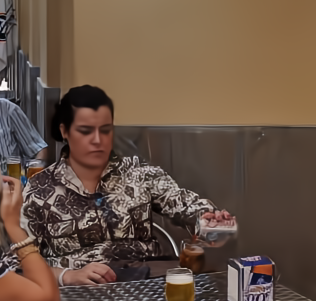}
        \caption*{Cheng20 (0.173/30.35/0.842)}
    \end{subfigure} 
    \\ 
    \begin{subfigure}{0.226\textwidth}
        \includegraphics[width=1.8cm]{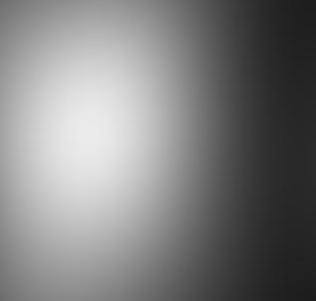}
        \includegraphics[width=1.8cm]{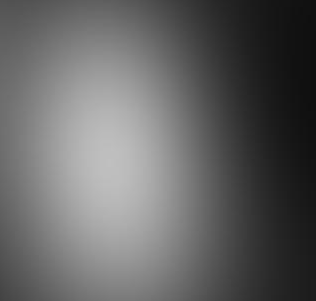}
        \caption*{Saliency maps}
    \end{subfigure}
    \hspace{0.005cm} 
    \begin{subfigure}{0.226\textwidth}
        \includegraphics[width=1.8cm]{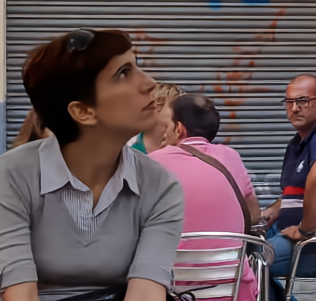}
        \includegraphics[width=1.8cm]{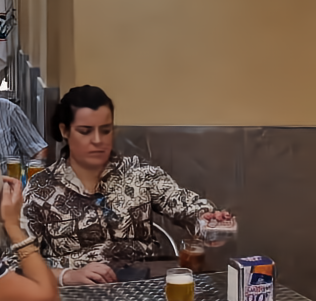}
        \caption*{Minnen18 (0.19/30.42/0.842)}
    \end{subfigure}
    \hspace{0.005cm} 
    \begin{subfigure}{0.226\textwidth}
        \includegraphics[width=1.8cm]{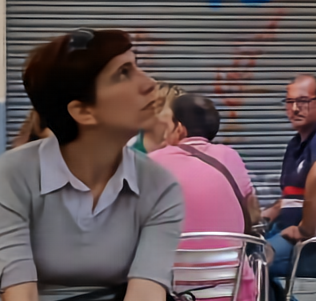}
        \includegraphics[width=1.8cm]{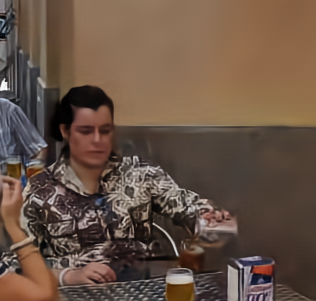}
        \caption*{Ball\'{e}18 (0.148/28.35/0.799)}
    \end{subfigure}
    \hspace{0.005cm} 
    \begin{subfigure}{0.226\textwidth}
        \includegraphics[width=1.8cm]{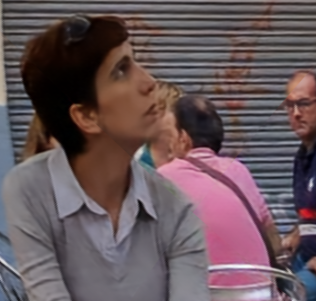}
        \includegraphics[width=1.8cm]{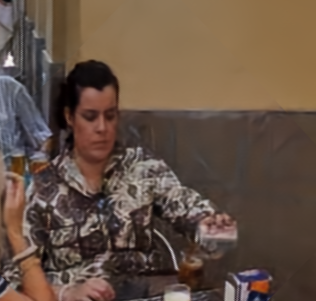}
        \caption*{OSLO (0.16/30.08/0.803)} 
    \end{subfigure}
    \caption{Visual comparison of two crops from “P26” of Salient 360! 2017 \cite{salient360_2017} test set decoded by different codecs around 0.15 bpp. The metrics under subfigures are (bpp↓ /SAL-PNSR↑ /WS-SSIM↑).}
    \label{fig:visual_results}
\end{figure*}

We evaluated the visual quality of decoded images from our method against other 2D and 360\degree\ image codecs using the Salient 360! 2017 dataset. As illustrated in \cref{fig:visual_results} with cropped sections of the P26 image, our method excels in preserving high-frequency details, particularly in areas like clothing, faces, and salient background details. Due to the computational cost of compressing 360\degree\ images at original resolution with LIC360 \cite{lic360}, refer to the supplementary materials for a detailed comparison. It's worth noting that while OSLO's metrics surpass Ball\'{e}18 in the spherical domain, its use of the Mollweide projection introduces visual distortions in ERP.

\subsection{Ablation Studies}
This section studies the effect of various components of the proposed architecture on overall performance.

\subsubsection{Effect of Fusion and Coordinate Refinement Network} We evaluated the Fusion network's impact by comparing the performance of our saliency model with and without it. The Base model, shown in \cref{tab:saliency_ablation_table}, uses pixel-wise multiplication of saliency maps from the Attention and Expert streams. Performance improvements are evident with the Fusion network, as seen in the table and visually in \cref{fig:saliency_ablation}. Pixel-wise multiplication often favors the Expert stream due to its tendency to highlight fewer salient regions, suppressing areas identified by the Attention stream. In contrast to ATSal \cite{dahou2020atsal}, the Fusion network merges maps using a learned method, ensuring a balanced representation from both streams.

\subsubsection{Effect of Coordinate Refinement Network} 
\begin{figure}[ht]
    \vspace{-20pt}
    \centering
    \begin{subfigure}[b]{0.32\linewidth}
        \centering
        \includegraphics[width=\linewidth]{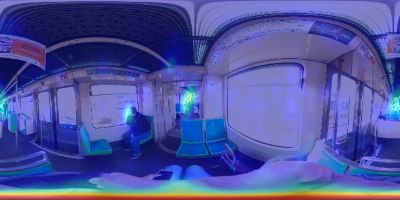}
        \caption{}
        \label{fig:short-a}
    \end{subfigure}
    \hspace{0.001cm}
    \begin{subfigure}[b]{0.32\linewidth}
        \centering
        \includegraphics[width=\linewidth]{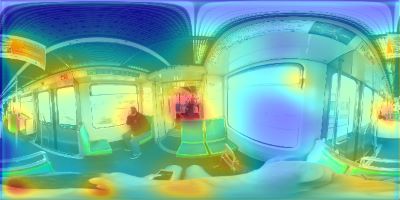}
        \caption{}
        \label{fig:short-b}
    \end{subfigure}
    \hspace{0.001cm}
    \begin{subfigure}[b]{0.32\linewidth}
        \centering
        \includegraphics[width=\linewidth]{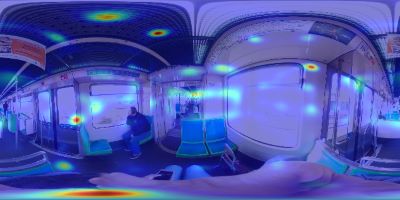}
        \caption{}
        \label{fig:short-c}
    \end{subfigure}
    \vfill
    \begin{subfigure}[b]{0.32\linewidth}
        \centering
        \includegraphics[width=\linewidth]{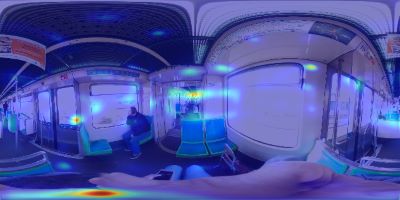}
        \caption{}
        \label{fig:short-d}
    \end{subfigure}
    \hspace{0.001cm}
    \begin{subfigure}[b]{0.32\linewidth}
        \centering
        \includegraphics[width=\linewidth]{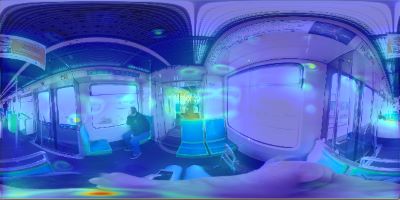}
        \caption{}
        \label{fig:short-e}
    \end{subfigure}
    \hspace{0.001cm}
    \begin{subfigure}[b]{0.32\linewidth}
        \centering
        \includegraphics[width=\linewidth]{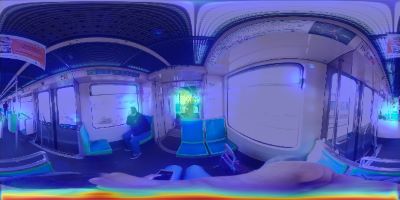}
        \caption{}
        \label{fig:short-f}
    \end{subfigure}
    \caption{ Visual comparison: (a) The ground truth saliency map. (b) The Attention stream result. (c) The Expert stream result. (d) Base (pixel-wise multiplication of (b) and (c)) result. (e) Base + Fusion result. (f) Base + Fusion + Coordinate Refinement result.}
    \label{fig:saliency_ablation}
    \vspace{-5pt}
\end{figure}

To evaluate the impact of the Coordinate Refinement network, we compare the performance of our saliency model with and without its inclusion. \cref{tab:saliency_ablation_table} shows that the best results in all saliency metrics are achieved when both the Fusion and Coordinate Refinement networks are employed. As it can be seen in the \cref{fig:short-f}, the inclusion of Coordinate Refinement networks results in a more detailed and accurate saliency map, similar to \cite{monroy2018salnet360}.
This approach avoids neglecting salient regions at the poles, contrasting with the traditional method of manually introducing an equator bias.

\newcolumntype{P}[1]{>{\centering\arraybackslash}p{#1}}
\newcolumntype{M}[1]{>{\centering\arraybackslash}m{#1}}
\begin{table}
  \centering
  \small 
  \setlength{\tabcolsep}{1.8pt} 
  \renewcommand{\arraystretch}{1.2} 
  \begin{tabular}{P{4.5cm} c c c c c}
    \toprule
    Method & KLD $\downarrow$ & CC $\uparrow$ & NSS $\uparrow$ & AUC-J $\uparrow$\\
    \midrule
    Base (pixel-wise multiplication) \cite{dahou2020atsal} & 0.822 & 0.618 & 0.708 & 0.651\\
    Base + Fusion & 0.639 & 0.652 & 0.810 & 0.697\\
    Above + Coordinate Refinement & \textbf{0.406} & \textbf{0.669} & \textbf{0.981} & \textbf{0.737}\\
    \bottomrule
  \end{tabular}
  \caption{Saliency ablation study on the test set of Salient360! 2017 Benchmark \cite{salient360_2017}.}
  \label{tab:saliency_ablation_table}
  \vspace{+16pt}
\end{table}

\subsubsection{Analysis of Latent Masking and Sal-MSE Loss} To assess the effects of latent masking and Sal-MSE loss, we trained our compression architecture with and without them, as shown in the top four rows of \cref{tab:compression_ablation}. While both methods individually enhance BD-PSNR and BD-rate performance as can be seen in rows 2 and 3 of \cref{tab:compression_ablation}, their combined application results in a BD-rate of -10.21\% and -12.85\% for WS-PSNR and SAL-PSNR, respectively. Notably, this combination boosts SAL-PSNR more than WS-PSNR, underscoring the significant benefits of their joint use with the training procedures described in \cref{sec:experimental}.

 
\begin{table}
    \centering
    \small
    \setlength{\tabcolsep}{2.5pt} 
    \begin{tabular}{c c c|c c}
        \hline
        \multirow{2}{*}{Method} & \multicolumn{2}{c|}{WS-PSNR} & \multicolumn{2}{c}{SAL-PSNR} \\
        & BD-psnr & BD-rate & BD-psnr & BD-rate \\
        \hline
        Base & 0 dB & 0 & 0 dB & 0 \\
        Base +Sal-MSE & 0.20 dB & -5.30\% & 0.26 dB & -6.30\% \\
        Base +Masking & 0.18 dB & -4.77\% & 0.29 dB & -7.11\% \\
        Base +Sal-MSE+Masking & \textbf{0.42 dB} & \textbf{-10.21\%} & \textbf{0.57 dB} & \textbf{-12.85\%} \\
        \hline
    \end{tabular}
    \caption{Compression ablation study on Salient360! 2017 test set using ELIC \cite{elic_comp} (Base) as baseline.}
    \label{tab:compression_ablation}
\end{table}

\section{Conclusion}

We introduce a saliency-aware end-to-end variable-rate model for 360\degree{} image compression, achieving SOTA results in ODI saliency detection and compression architectures. Our learned model tackles oversampling and equator bias in 360\degree{} ERP images by capturing and fusing local and global saliency. Integrating Sal-MSE loss with latent masking and applying data augmentation to address oversampling enhances visual quality in salient regions. Our method shows superior WS-PSNR performance and excels notably in the SAL-PSNR measure. Our saliency-aware model underscores the role of visual attention in ODI image compression, offering significant BD-rate and BD-PSNR savings.

\bibliographystyle{IEEEbib}
\bibliography{strings,refs}

\end{document}